\documentclass[twocolumn,showpacs,showkeys,preprintnumbers,superscriptaddress,amsmath,floatfix,amssymb,secnumarabic]{revtex4}
\usepackage[colorlinks=true]{hyperref}
\usepackage{graphicx}
\usepackage{epsfig}
\usepackage{slashbox}

\newcommand{\beq}{\begin{equation}}
\newcommand{\eeq}{\end{equation}}

\def\dalam{\hbox
{\vrule\vbox{\hrule\hbox to 1ex{ \hfill}\kern 1 ex\hrule}\vrule}}
\def\sign{\hbox{sign}}
\def\1/2{\hbox{$ {1 \over 2}$ }}

\def\h{\hbar}
\def\i/h{{i \over \h}}

\def\inf{\infty}
\def\pd{\partial} 
\def\v{\vec}

\def\a{\alpha}

\def\g{\gamma} \def\G{\Gamma} 
\def\d{\delta} \def\D{\Delta}
\def\l{\lambda} 
 
 \def\tE{\tilde {E}}

\def\s{\sigma}\def\S{\Sigma}
\def\r{\rho} 
\def\x{\xi}
\def\c{\chi} 
\def\vf{\varphi}
\def\f{\phi} \def\F{\Phi}
 
\def\p{\psi} \def\P{\Psi}

\def\W{\Omega}
\def\tt{\theta}

\def\<{\langle}
\def\>{\rangle}

\def\({\left(}
\def\[{\left[}
\def\){\right)}
\def\]{\right]}

\begin{document}
\sloppy

\title{Spontaneous spherical symmetry breaking in atomic confinement}

\author{K.~Sveshnikov}
\email{costa@bog.msu.ru} \affiliation{Department of Physics and
Institute of Theoretical Problems of MicroWorld, Moscow State
University, 119991, Leninsky Gory, Moscow, Russia}

\author{A.~Tolokonnikov}
\email{Tolokonnikov@physics.msu.ru} \affiliation{Department of Physics and
Institute of Theoretical Problems of MicroWorld, Moscow State
University, 119991, Leninsky Gory, Moscow, Russia}

\date{July 13, 2016}

\begin{abstract}
The effect  of spontaneous breaking of initial $SO(3)$ symmetry is shown to be possible for an H-like atom in the ground state, when it is confined in a spherical box under general boundary conditions of ``not going out''  through the box surface (i.e. third kind or Robin's ones), for a wide range of physically
reasonable values of system parameters. The reason is that such boundary conditions could yield a large magnitude of electronic wavefunction in some sector of the box boundary,  what in turn promotes atomic displacement from the box center towards this part of the boundary, and so the underlying $SO(3)$ symmetry spontaneously breaks. The emerging Goldstone modes,  coinciding  with rotations around the box center, restore the symmetry by spreading  the atom over a spherical shell localized at some distances from the box center. Atomic confinement inside the cavity proceeds dynamically --- due to the  boundary condition  the deformation of electronic wavefunction near the boundary works as a spring, that returns the atomic nuclei back into the box volume.

\end{abstract}

\pacs{31.15.A-, 32.30.-r, 37.30.+i}
\keywords{confined quantum systems, broken symmetries, H-like atom, Wigner-Seitz model }

\maketitle

\subsection*{1. Introduction}

The concept of broken symmetries is a cornerstone of such important physical effects as the Higgs mechanism in the electroweak sector of the standard model, confirmed recently by the discovery of  corresponding boson, ferromagnetism, superconductivity and superfluidity. The common feature of all these phenomena is that they proceed in systems with infinite degrees of freedom like a phase transition with rise of nonzero quasi-average. As a result, the initial symmetry of the system reduces to a subgroup and an order parameter appears, which as a rule coincides with nonvanishing quasi-average. In systems with finite degrees of freedom, where  field-theoretic/statistical nature is absent in principle,  symmetry breaking doesn't reveal such direct analogy with a phase transition. Nevertheless, in this case all the main features including nonzero quasi-average as well as Goldstone modes (in the case of broken continuous symmetry), which give rise to significant changes in the system properties,  should exist at the same right.

The purpose of this letter is to explore such an effect, which takes place by confinement of atomic systems in a closed simply connected cavity. Such systems attract now considerable amount of theoretical and experimental activity \cite{Jask}-\cite{Dyson}, \cite{Koo}-\cite{Wiese1}.  So far, starting from the works of Michels and de Boer \cite{Michels}, Sommerfeld and Welker \cite{Sommerfeld}, main attention has been devoted to the properties of atoms and molecules, confined by an impenetrable or partially
penetrable potential wall (\cite{Jask}-\cite{Sen0},\cite{Koo}-\cite{Aquino1} and refs. therein).  However,   actually  general boundary conditions of ``not going out''   imply a quite different picture, where the particle WF doesn't unavoidably vanish at the box boundary  (\cite{Pupyshev}-\cite{Wiese1} and refs. therein). As a consequence, the lowest energy levels of confined atomic system undergo a deep reconstruction, that might cause a spontaneous breakdown of the initial symmetry of the system. In particular, it occurs for atomic H placed in a spherical cavity, when for a wide range of physically reasonable values of system parameters a nonzero shift of  atomic nuclei (the proton) from the center of cavity in the atomic ground state takes place, leading to broken rotational symmetry. In accordance with general features of broken symmetries, the shift is accompanied by emergence of corresponding Goldstone modes,  representing  fluctuations  of spontaneous average under group transformations, in this case $SO(3)$. In  turn, these modes coincide with rotations around  the cavity center and  so restore the original symmetry, while  atomic states acquire rotational quantum numbers and some additional nontrivial properties. Preliminary, but important results on this subject, motivated by the study of an endohedral atom in a fullerene cage \cite{Dolmatov},\cite{Connerade1},\cite{Connerade2}, have been reported  in ref. \cite{ZRP} a decade ago, where a simplified semi-analytic model, based on zero-range potential technique to mimick the interaction of an active electron with the residual atomic  core, was considered, demonstrating the possibility of spontaneous $SO(3)$ breakdown for strong attractive interaction. Here we present another approach to the problem starting from general ``not going out'' conditions, which allows for a detailed (quasi)exact study of the effect of rotational symmetry  breaking not only for attractive, but for repulsive interaction too, as well as for various nontrivial asymptotical regimes.

It is worth-while to note, that  the  general boundary conditions of ``not going out''  don't unavoidably imply  genuine trapping of a particle by
a cavity, rather they could be caused by a wide range of reasons, as in  the Wigner-Seitz model of alkaline metal \cite{WS}, when the particle state is delocalized from the beginning.
 The latter circumstance turns out to be quite important,
since in some cases the cavities, where a
particle or an atom could reside, form a lattice, similar to that of an alkaline metal,  like
certain interstitial sites of a metal supercell, e.g. next-to-nearest octahedral
positions of palladium fcc lattice \cite{Alefeld}-\cite{PdHx}. In this case a particle  (or valence atomic electron, provided that the whole lattice of cavities is occupied by atoms) finds itself in a periodic potential of a cubic lattice, and so on the boundary of corresponding Wigner-Seitz cell its ground state WF should be subject of Neumann condition (\ref{f5}), what is a special case of general ``not going out'' problem.

\subsection*{2. General treatment  of a ``not going out'' state}

The general approach to description  of a ``not going out'' state for a particle  in a vacuum
cavity $\W$ with boundary $\S$ starts with  the following energy functional  \cite{Pupyshev}-\cite{Pupyshev1}
 \beq \label{f1} E[\p]=\int_\W \! d \vec r \
\left[ {\h^2 \over 2m } | {\vec \nabla } \p|^2 + U(\vec r ) \
|\p|^2 \right] +
\nonumber \\ \eeq
\beq \label{f1}
 + \ {\h^2 \over 2m} \int_\S \! d\s \ \l (\vec r) \
|\p|^2  \ , \eeq
 where $U(\vec r)$ is the potential   inside  $\W$, while the surface term $\int_\S$ corresponds to  contact interaction of particle with medium, in which the cavity has been  formed, on its boundary. The properties of this surface interaction are given by a real-valued function $\l (\vec r)$. A more realistic model of interaction with environment should imply the boundary in the form of a potential shell with definite magnitude and size \cite{Dolmatov},\cite{Connerade1},\cite{Connerade2},\cite{Sv1}, but for our purposes it doesn't play any significant role, since all the main effects of broken spherical symmetry show up already by contact interaction in (\ref{f1}).

From the variational principle with normalization condition  $\<
\p|\p \>=\int_\W \! d \vec r \  |\p|^2 =1$ one obtains \beq
\label{f2}
 \left[ -{\h^2 \over 2m } \D + U(\vec r )  \right] \p=E \p
\eeq inside $\W$ combined with Robin's (or third kind) boundary condition imposed on $\p$
on the surface  $\S$ \beq \label{f3} \left. \[ \vec n \vec \nabla +
\l(\vec r )  \] \p \right|_\S=0 \ , \eeq with $\vec n $ being the
outward normal to $\S$. The  ``not going out''  property is fulfilled
here via vanishing normal to $\S$ component of the  quantum-mechanical flux
 \beq \label{f5} \vec j = {\h \over 2m i} \ \(
\p^{\ast} \vec \nabla  \p - \p \vec \nabla \p^{\ast} \) \  \eeq
at the box boundary $ \left.  \vec n \vec j \right|_\S=0
\ . $ At the same time, tangential components of  $\vec j$ could be
remarkably different from zero on $\S$ and so the particle could
be found quite close to the boundary with a marked probability.
Such a picture is similar to that of the Thomas-Fermi model
of many-electron atom \cite{LL}, as well as to quark bag models of hadron
physics \cite{MIT bag},\cite{Chiral bag}.

For a spherical cavity the spectral
problem (\ref{f2}-\ref{f3}) is self-adjoint and so contains all the
required properties for a correct quantum-mechanical description of
a non-relativistic particle confined in $\W$. For a more complicated geometry the set of requirements to $\W$ and $\S$, under which the problem (\ref{f2}-\ref{f3}) allows for a self-consistent treatment, is discussed in \cite{Sen2},\cite{Wiese},\cite{Pupyshev1}.

When $\l=0$, interaction of the particle with environment is
absent and so eq. (\ref{f3}) transforms into  Neumann (second kind)
condition \beq \label{f5} \left.  \vec n \vec \nabla \p
\right|_\S=0 \ , \eeq what is similar to the boundary condition
of confinement for a scalar field in relativistic bag models \cite{MIT bag}.
Moreover, condition (\ref{f5}) appears in the Wigner-Seitz model of
an alkaline metal \cite{WS} and describes an opposite picture, namely delocalization of valence
electrons creating the metallic bond, by continuing the electronic
WF periodically  in the lattice. So the ``not going out'' state with Neumann condition turns out to be of special interest, since in media with long-range order such vacuum cavities could  form a (sub)lattice \cite{Alefeld}-\cite{PdHx}. Let us also mention, that in the case of atomic H the condition (\ref{f5}) is nothing else, but the boundary condition for the Wigner-Seitz cell in hypothetical metallic atomic phase (see e.g. \cite{Max} and refs. therein), which should be the simplest alkaline metal.

If $\l \to \infty$, then   (\ref{f3}) turns into Dirichlet
condition \beq \label{f6} \left.  \p \right|_\S=0 \ , \eeq and so
describes  confinement by an impenetrable barrier. However, when $\l \to - \infty$, the answer depends on the size of cavity. When the latter is finite, the atomic position in the center of cavity becomes unstable and the atom sticks to the boundary, whereas if the cavity size grows infinitely, the curvature of the boundary becomes negligibly small and there appears a separate nontrivial problem of atomic states over a plane with boundary condition (\ref{f3}), whose solution depends strongly on concrete relation between $\l$ and atomic nuclei charge  $q$ (for a more detailed discussion of the latter problem see Section 4).

\subsection*{3. Atomic H in the center of cavity}

The problem under consideration concerns  atomic H with nuclei charge
$q$ in a spherical cavity with size $R$ and boundary condition (\ref{f3}), when the surface interaction is given by a constant $\l$ and so the system reveals spherical symmetry. Besides \cite{ZRP},  most of works on this subject   starting from  \cite{Michels}, \cite{Sommerfeld} till nowdays \cite{Jask}-\cite{Sen0},\cite{Koo}-\cite{Wiese1},  propose that the atomic nuclei should reside in the center of cavity. Then  spherical symmetry is maintained
and  the  radial electronic WF  with orbital momentum $l$
up to a numerical factor takes the following form \cite{LL}
\beq \label{f7} R_l (r) =e^{-\g r} r^l \ \F (b_l , c_l , 2 \g r) \ , \eeq
where
\beq \label{f8} \g=\sqrt{-2E} \ ,
\ b_l=l+1-q/\g \ , \ c_l=2l+2 \ , \eeq and  $\F(b,c,z)$ is  the first kind
confluent hypergeometric (Kummer) function. Definition, notations and main properties of the Kummer
function follow ref. \cite{Beytmen}. In what follows, in order to
provide an effective comparison of  results obtained previously, universal relativistic units $\h=c=1$ are used, wavenumber and energy are expressed in units of the electron mass
$m$, while distances --- in units of corresponding Compton length
$1/m$, and e-m coupling constant $\a$ is included in the nuclei charge $q=Z \a$. Such units are in particular quite convenient for  accounting of relativistic effects as well as for numerical calculations in the Schroedinger case, since  step one in  Compton length units corresponds to the continuous limit of nonrelativistic problem.

Substituting (\ref{f7}) into the boundary
condition  (\ref{f3}) yields the following equation for  energy
levels
\beq \label{f9}
\[ q/\g + (\l-\g)R -1 \]  \F_R + \[l+1 -q/\g\] \F_R(b+) =0 \ ,
\eeq
where
\beq \label{f10} \F_R=\F (b_l , c_l , 2 \g R) \ , \
\F_R(b+)=\F (b_l+1 , c_l , 2 \g R) \ . \eeq

There are the following properties of electronic levels for an H-like atom in the center of cavity, that are of interest  for further analysis. The most significant
changes in the spectrum take place for $R \to 0$. Here it should be noted, that for  atomic H the limit $R \to 0$ requires some care, since  relativistic effects give rise to the restriction $R \geq 10$ for the cavity sizes, where such an approach to the confinement problem, based on boundary condition (\ref{f3}), should be valid \cite{Sv1}. So in what follows the limit $R \to 0$ should be understood either as a purely mathematical property of equations under consideration, or as decreasing $R$ up to $R \sim 10$.

There are two  types of
lowest levels for atomic H in dependence on
relation between $\l$ and $q$. The first one takes place under assumption, that for
$R \to 0$ the wavenumber $\g$ remains finite, and so in the
vicinity of $R=0$ it could be represented by a series
\beq \label{f11} \g(R)=\g_0+ \g_1 R + \g_2 R^2 + ... \ . \eeq Expanding
$\F_R \ , \F_R(b+)$
as a power series in  $R$, to the lowest order one obtains from (\ref{f9})
that $l=0$, and by proceeding further
\beq \label{f12} \l=q \ ,
\quad \g_0^2=q^2 \ , \quad \g_n=0, \quad  n \geq 1  \ . \eeq
From (\ref{f12}) there follows the result, already  mentioned in \cite{Pupyshev},\cite{Sen2}, that when $\l=q$, then the ground state
energy of  atomic H  in a cavity for any $0 < R \leq \inf $
precisely coincides with that of $1s$-level of the free atom \beq
\label{f13} E_{ground}(R) = E_{1s}=-q^2/2 \  , \eeq
while the electronic WF coincides inside the  cavity with corresponding one of free H.

Another type of levels
is found by assumption, that in the vicinity of
$R=0$ the wavenumber $\g$ is represented by a series \beq
\label{f14} \g(R)= {\x \over \sqrt{R}} + \x_0 + \x_1 \sqrt{R} +
\ldots \  . \eeq Substituting (\ref{f14}) into eq.(\ref{f9}), to the
lowest order in  $\sqrt{R}$ one obtains again $l=0$, while  higher
orders of expansion in   $\sqrt{R}$ yield \beq \label{f15}
\x^2=3(q-\l) \ , \quad \x_0=0 \ , \quad \x_1={q^2 + 3q\l +6\l^2
\over 20\x} \ , \quad \ldots \ . \eeq
As  a result, for such  type of  $s$-levels  in a
cavity one obtains the following dependence on the cavity size for
$R \to 0$
\beq \label{f16} E_{ground}(R) \to - { 3(q-\l) \over 2 R}
- {q^2 + 3q\l +6\l^2  \over 20} + O(\sqrt{R}) \ ,
\nonumber \\ \eeq
\beq \label{f16} R  \to 0  \ , \eeq
what coincides with  (\ref{f13}) for $\l=q$.

Qualitative explanation of linear dependence on $q$
and  $\l$ in (\ref{f16}) is quite simple. As for a particle in a spherical well \cite{Wiese}-\cite{Sv1},
for $R \to 0$ the electronic WF of such  $1s$-level inside a cavity
becomes almost constant, and substitution of corresponding normalization constant into (\ref{f1}) yields immediately the leading term in the asymptotics (\ref{f16}).

The peculiar features of the problem (\ref{f2}-\ref{f3}) show up for $R \to \inf$ as well, when it could be found via asymptotic expansion for $\F_R \ , \
\F_R(b+)$ in (\ref{f9}), that in the case of surface attraction
$\l <0$ there exists  besides the discrete
spectrum of the free atom one more level $\tE (R)$  with negative  limiting value $\tE(\inf) = -
\l^2/2$  and  power behavior for   $R \to \inf$
\beq
\label{f17} \tE (R) \to - \l^2/2 -  (q-\l)/R + O(1/R^2) \ , \ R \to
\inf \ . \eeq
For  $\l < -q < 0 $ such a power level $\tE
(R)$ turns out to be the lowest electronic $s$-level for all $R$ and looks like a
shifted downwards hyperbole, as for a particle in a well \cite{Pupyshev},\cite{Sv1}.

 At the
same time,  the levels originating from the discrete spectrum of free H, tend for $R \to \inf$  to their asymptotical values, corresponding
to those of free H, exponentially fast  \cite{Pupyshev},\cite{Sen2},\cite{Sv1}. Moreover, for atomic H in a spherical box the Runge-Lenz vector isn't conserved yet \cite{Wiese},\cite{Wiese1}, hence  these levels should be labeled with two quantum numbers $n=n_r+1$ and $l$. In particular, for the $ns$-levels one finds
 \beq \label{f18}
E_n(R) - E_n \to  \[ {\g_n \over n! } \]^2 \ { \l -\g_n \over \l
+\g_n} \ \( 2 \g_n R\)^{2n} \ e^{-2\g_n R} \ ,
\nonumber \\ \eeq
\beq \label{f18} \g_n R \gg 1 \ , \eeq
where \beq \label{f19}
E_n= - \g_n^2 / 2 \ , \quad \g_n = q / n  \ , \quad n=1,2,\dots \ , \eeq
are the electronic $ns$-levels of the free atom. Remark, that levels with $\g_n < \l$ should approach their asymptotics from above, while those with $\g_n > \l$ from below.

It should be specially  noted, that the asymptotics (\ref{f18})
turns out to be an exceptional feature of those confined atom
electronic levels, which originate from the discrete spectrum of the free
atom, since such asymptotics is created  by approaching the
argument of the factor $\G^{-1}(b)$, entering the asymptotics  of
the Kummer function $\F (b,c,z)$, to the pole $b \to -n_r, \
n_r=0,1,\dots \ .$ Asymptotics for $R \to \inf$ of all the other
electronic levels in a cavity, which originate from the continuous
spectrum of the free atom, and of the additional powerlike level (\ref{f17}), caused
by attractive interaction with environment, turns out to be  a
power series in $1/R$, and their  asymptotical values could be
either non-negative only, or for $\l <0$ contain one negative
point $\tE (\inf)=-\l^2/2$ \cite{Sv1}.

For  $\l = \pm \g_n$ the asymptotics (\ref{f18})  modifies in the
next way. The exponential behavior is preserved, while
the non-exponential factor undergoes changes in such a way, that
the $ns$-levels approach their asymptotics of the free atom from above
only. For  $\l =\g_n >0$ their asymptotics takes the form
 \beq \label{f20}
E_n(R) - E_n \to  (n-1) \[ {\g_n \over n! } \]^2  \ \( 2 \g_n
R\)^{2(n-1)} \ e^{-2\g_n R} \ ,
\nonumber \\ \eeq
\beq \label{f20} \g_n R \gg 1 \ , \eeq
while for the lowest level  $E_1(R)$ the exponential part disappears completely, since in this case $\l=\g_1=q$, and as it was mentioned above,  $E_1(R)$ becomes a constant, which coincides with  $E_{1s}=-q^2/2$.

For $\l =-\g_n < 0$ instead of (\ref{f18}) one obtains
 \beq \label{f21}
E_n(R) - E_n \to { 1 \over n+1} \  \[ {\g_n \over n! } \]^2  \ \(
2 \g_n R\)^{2(n+1)} \ e^{-2\g_n R} \ ,
\nonumber \\ \eeq
\beq \label{f21} \g_n R \gg 1 \ , \eeq and
moreover, the limiting point $\tE (\inf)$ of the level $\tE (R)$ with the power
asymptotics (\ref{f17})  coincides with the corresponding level $E_n$ of the
free atom (\ref{f19}), what in turn represents a remarkable example
of von Neumann-Wigner avoiding crossing effect, i.e. near levels reflection  under perturbation \cite{LL},\cite{NW}
--- infinitely close to each other for $R \to \inf$ levels
$E_n(R)$ and  $\tE (R)$  should for decreasing $R$ diverge in
opposite directions from their common limiting point  $E_n$.
Perturbation in this case is performed by the atomic nuclei
Coulomb field, since  under general boundary conditions (\ref{f3}) the electronic WF doesn't vanish on the
cavity boundary, and so  for $R \gg 1$ the maximum of electronic
density should be shifted into the region of large distances
between the electron and nuclei, where the contribution of the
Coulomb field is negligible compared to boundary effects. When
$R$ decreases, the Coulomb field increases, hence   $E_n(R)$
should go upwards according to (\ref{f21}), while $\tE (R)$ goes
downwards according to the asymptotics \beq \label{f22} \tE (R) \to
E_n -  {n+1 \over n} \ {q \over R} + O(1/R^2) \ , \ R \to \inf \ .
\eeq

 So the lowest electronic level of atomic H, confined in a cavity with Robin's condition (\ref{f3}), turns out to be the following \cite{Pupyshev},\cite{Sen2},\cite{Sv1}. For $\l=q$ it acquires the constant value $E_{1s}$ of the free atom, for $\l>-q$  behaves for  $R \to 0$ according to (\ref{f16}) with an energy shift depending on $\sign \(\l -q\)$ and for  $R \to \inf$ it approaches $E_{1s}$ exponentially fast, while for $\l \leq -q <0$  transforms into the level  $\tE(R)$ with power asymptotics (\ref{f17}).

\subsection*{4.  The machinery of spherical symmetry breaking for atomic ``not going out'' state}

Actually the assumption of spherical symmetry of the atomic ground state in a spherical box doesn't hold for $\l < q$. The reason is that for boundary conditions (\ref{f3}) electronic WF could be localized in the vicinity of the cavity boundary, hence  the most favorable atomic configuration should be that one, when the atomic nuclei is shifted from the center of  cavity. What is here the most nontrivial, that while for $\l <0$ the atomic displacement turns out to be an almost obvious consequence of electron attraction to the cavity boundary, for $\l >0$, i.e. for electron reflection from the boundary,  the emerging asymmetric distortion of  electronic WF  could yield an energy decrease as well. Moreover, there appears a quite complicated dependence on the cavity size. The crucial role in this effect is played by the boundary condition (\ref{f3}). The fine structure and other spin effects turn out to be negligibly small compared to those coming from  the boundary condition and so are neglected in what follows.

As a first step let us consider this effect via special kind trial function. For these purposes the coordinate frame with origin in the cavity center turns out to be the most convenient. Assuming that the adiabatic approximation is valid,  for atomic nuclei  placed in the point  $ \v a=(0,0,a)$ the electronic  hamiltonian takes the form (in what follows the genuine atomic H with $Z=1$ is considered, hence $q=\a$)
\beq \label{f23}
H_{el}=- \1/2 \D_{\v r} -  {\a \over |\vec r - \vec a|} \ .
\eeq
The trial function is chosen as a superposition of N first angular harmonics with zero momentum projection on z-axis (rotation over z-axis maintains the symmetry, so $l_z$ remains an actual quantum number, while the energy eigenvalue is minimized for $l_z=0$)
\beq \label{f24}
\P_{tr}(r,\tt)=\sum_{l=0}^N  c_l \ R_l (r) \ P_l (\cos \tt ) \ ,
\eeq
and with radial functions of the structure similar to that of lowest H-levels with momentum $l$
\beq \label{f25}
R_l(r)=d_l \ r^l e^{-\g_l r} .
\eeq
The variation parameters are here the coefficients $c_l$, which could be taken real from the beginning, while $\g_l \ , d_l$ are determined via boundary condition and normalization  of $\P_{tr}$
\beq \label{f26}
\g_l=\l + {l \over R} \ , \quad d_l= \( { 2l+1 \over 4 \pi } \ { (2 \g_l)^{2l+3} \over \G (2l+3,0,2\g_l R) } \)^{1/2} \ ,
\eeq
with $\G(z, x_0, x_1)=\int \nolimits_{x_0}^{x_1} \! t^{z-1} e^{-t} dt$ being the generalized incomplete gamma-function. With $\g_l$ taken in the form (\ref{f26}) such a trial function  fulfills  exactly the boundary condition (\ref{f3}), and so a nonzero  displacement shows up already for $\l>0$.

The normalization of the trial function takes the form
\beq \label{f27}
\< \P_{tr} | \P_{tr} \>= \sum_{l=0}^N \ c_l^2 \ ,
\eeq
while for the energy functional (\ref{f1}) one obtains the following quadratic form in $c_l$
\beq \label{f28}
E[\P_{tr}]=\< \P_{tr} |H| \P_{tr} \>= \sum_{l=0}^N \ c_l A_{ls} c_s \ ,
\eeq
where
\beq \label{f29}
A_{ls}=(K_l+ \l \ B_l) \d_{ls} - 2 \a V_{ls} \ .
\eeq
In the matrix $A_{ls}$ the diagonal kinetic $K_l$ and surface $B_l$ terms can be written as
\beq \label{f30}
K_l={ d_l^2 \over (2l+1) (2\g_l)^{2l+1} }
 \left[ {1 \over 4} \G (2l+3,0,2\g_l R) \ -  \right.
\nonumber \\ \eeq
\beq \label{f30}
\left. - \ l \G (2l+2,0,2\g_l R) + (2l+1) \G (2l+1,0,2\g_l R) \right] \ ,
\eeq
\beq \label{f31}
B_l={ d_l^2 \over (2l+1) } \ R^{2 l+2} e^{-2 \g_l R} \ ,
\eeq
while for the Coulomb interaction the terms $V_{ls}$
\beq \label{f32}
V_{ls}=\sum_{k=|l-s|}^{l+s} W_{lsk}(a) \left( l \ s \ k \atop 0 \ 0 \ 0 \right)^2 ,
\eeq
where
\beq \label{f33}
W_{lsk}(a)=d_l \   \left(  { \G (l+s+k+3,0,(\g_l+\g_s) a) \over a^{k+1} (\g_l+\g_s)^{l+s+k+3} } \ + \right.
\nonumber \\ \eeq
\beq \label{f33}
\left.  + \ a^k  { \G (l+s-k+2,(\g_l+\g_s) a,(\g_l+\g_s) R) \over  (\g_l+\g_s)^{l+s-k+2} }  \right) \ d_s \ ,
\eeq
 are responding.

$E_{tr}$ is determined from the variational problem for  $c_l$ via secular equation
\beq \label{f34}
\det |A_{ls}- E_{tr}\d_{ls}|=0 \ ,
\eeq
which could be easily solved by means of standard numerical recipes. The resulting dependence $E_{tr}(a)$, obtained for $N=10$, is shown on Fig.1
\begin{figure}
\includegraphics[width=8.6 cm]{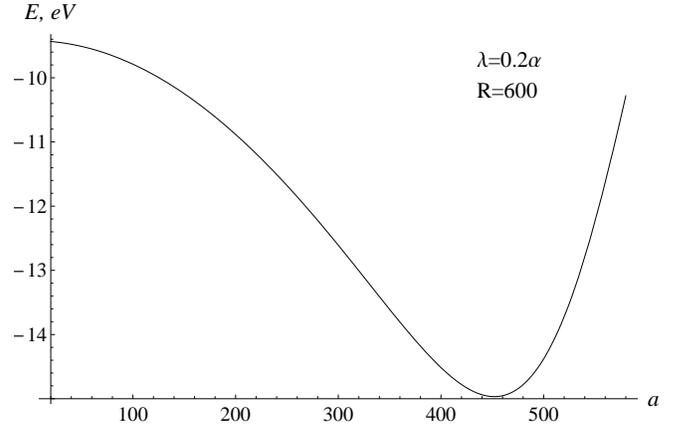}\\
\caption {$E_{tr}(a) $ for $R=600, \ \l=0.2 \a$.}
\label{fig:E_of_a_sample_var}
\end{figure}
The minimum on the curve $E_{tr}(a)$ depends strongly on actual values of $\l$ and $R$. Tab.1 shows the relation between the energy $E_0(0)$ of unshifted H lowest level, calculated from (\ref{f9}), and the minimal value $E_{tr}(a_{min})$ found for $E_{tr}(a)$.

\beq
{\small
 \begin{tabular}{c|c|c|c|c|c|c|c}
\backslashbox{$\lambda /\alpha$}{$R \ \ $} & 200 & 400 & 600 & 800 & 1000 & 1200 & 1400 \\
 \hline
0.5 & -0.473 & -0.388 & -1.669 & -2.839 & -3.279 & -3.362 &  -3.362  \\
0.0 &  1.425 &  4.673 &  4.005 &  2.169 &  0.512 & -0.843 &  -1.963 \\
-0.5 & 4.530  & 11.312  & 12.968 & 12.078 & 9.977 & 8.086 & 6.505  \\
-1.0 & 8.001 &  16.233  & 17.719 & 17.241  & 16.213  & 15.082 & 13.998 \\
-1.5 & 11.221  & 19.188   & 19.890 &  18.887 & 17.494 & 16.1 & 14.822
\end{tabular} }
\notag \eeq
{\small Tab.1. The values of $\D E=E_0(0)-E_{tr}(a_{min})$, given in eV, for $R=200-1400$ and 5 values  of $\l/\a$ from the range  $-1.5\leq \l/\a \leq 0.5$. }\vskip 0.3 cm

So in the trial function approximation  (\ref{f24})  there is no effect for $\l/\a=0.5$ and any $R$ from the range considered above, while for other $\l$ the shift is already present, but the dependence of the depth of shifted minimum on $R$ isn't monotonic --- the mostly pronounced effect is achieved for $400 < R < 600$.

(Quasi)exact analysis of atomic displacement is performed numerically on the basis of the following algorithm. Let us pass to the coordinate frame with origin in the proton. Assuming that the proton shift from cavity center is  $a$ in Oz direction, for the distance $r_{\S}$  between the proton and cavity boundary one finds the following dependence on the polar angle $\tt$
\beq \label{f35}
 r_\S (\tt)= \sqrt { R^2 - a^2 \sin^2 \tt  } - a \cos \tt \ ,
\eeq
operator $\left. \vec n \vec \nabla \right|_\S $ in the proton rest frame takes the form
\beq \label{f36}
\left.  \vec n \vec \nabla \right|_\S=\left. \( A \ { \pd \over \pd r } + B \ { \pd \over \pd \tt } \) \right|_{r=r_\S} \ ,
\eeq
where
\beq \label{f37}
A={  \sqrt { R^2 - a^2 \sin^2 \tt  } \over R } \ , \quad B=- { a \sin \tt \over R \ r_\S (\tt) } \ .
\eeq
The ground state electronic WF is assumed to be of the form (\ref{f24}) with $R_l(r)$ being now the radial Coulomb functions (\ref{f7}-\ref{f8}), while coefficients $c_l$ are determined from the boundary condition (\ref{f3}), what leads to secular equation for the energy levels of the form
 \beq \label{f38}
\det I=0 \ ,
\eeq
where
 \beq \label{f39}
I_{ls}= \int_{-1}^1 \! dx \[ \( A R'_l + \l R_l \) P_l (x)  + B R_l P_l^1 (x) \]_{r=r_\S} P_s (x) \ .
\eeq
In (\ref{f39}) the argument of Kummer function is $ r_\S$ (\ref{f35}), where $\cos \tt =x$, thence the integral in (\ref{f39}) cannot be calculated analytically, but allows for a detailed numerical analysis. For  $100 < R < 1000$ the precision of order $0,01$ eV for the lowest level is achieved for $N \geq 12$. Check-up of calculations based on (\ref{f38}-\ref{f39}) is performed by means of direct solution of initial Schroedinger eq. with hamiltonian (\ref{f23}) via shooting from the cavity center into boundary condition with such a number of harmonics, that provides relative error to be not more than $5 \%$. For these purposes  18-36 angular harmonics are used, while for calculations based on  (\ref{f38}-\ref{f39}) their number is 12-16, but in the latter case there appears an additional problem with controlling the required precision in (\ref{f39}), since it contains integration of oscillating functions with very large, exceeding several orders in magnitude, jumps of amplitudes, what requires application of arbitrary-precision (bignum) arithmetics.

 The typical behavior of the lowest electronic level $E_0(a)$ compared to the estimate via trial function (\ref{f24}) is shown on Fig.2. On the whole, it reproduces the result obtained by means of (\ref{f24}), but now the displacement effect takes place for sufficiently more large $\l >0$, while the well, where  the energy is minimized, turns out to be much deeper with a pronounced barrier for $a \to R$, which emerges due to significant increase of deformation, hence of gradients, in electronic WF, when the proton  approaches the box boundary.

 \begin{figure}
  \includegraphics[width=8.6 cm]{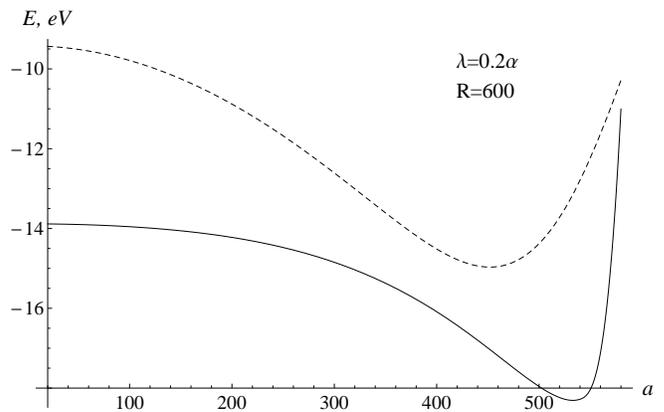}
\caption {$E_0(a) $ (solid line), $E_{tr}(a) $ (dashed line) for $R=600, \ \l=0.2 \a$.}
\label{fig:E_of_a_sample}
\end{figure}
\vskip 0.3 cm

 Phase diagram for regions of existence and absence  of the displacement effect in parametrization $\l/\a \ ,R $ is shown on  Fig.3. The boundary curve between these regions is determined from relation $\pd E_0 (a)/\pd a =0$ for $a \to 0$ and henceforth is marked as $\l^{\star}(R)$ or $R^{\star}(\l)$. The displacement effect is always absent for $\l \geq \a$, what is a direct consequence of that for the lowest electronic state in a cavity with boundary condition (\ref{f3}) for  $\l=\a$  and any $R$ there exists the exact solution  (\ref{f13}) with H placed in the center, which coincides with 1s-level of the free atom. For larger $\l$ the reflection from the boundary becomes even greater uniformly in all directions, hence H remains in the cavity center. For $\l < \a$ the atomic displacement depends on relation between $\l$ and $\l^{\star}(R)$. When  $\l<\l^{\star}(R)$, a nonzero displacement becomes an immanent feature of  atomic ground state, the more pronounced, the smaller the ratio $\l/\l^{\star}$ turns out to be.

  \begin{figure}
  \includegraphics[width=8.6 cm]{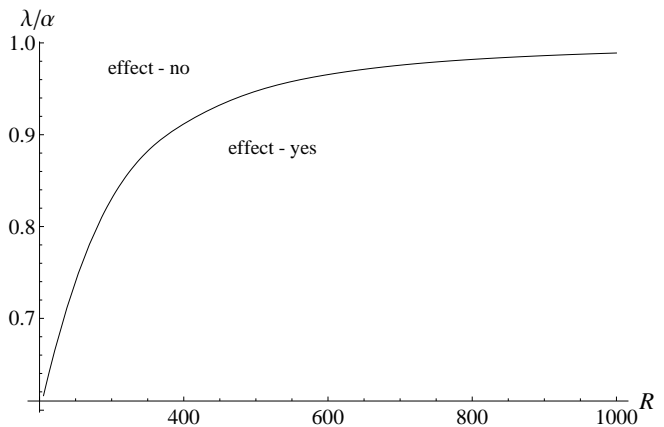}
\caption {Phase diagram for the displacement effect.}
\label{fig:lambda_of_R}
\end{figure}

 The dependence of the shifted electronic energy minimum  $E_{min}=E_0(a_{min})$ on $\l/\a$ for cavity sizes from the range $100 < R < 1000 $ is shown on Fig.4. Fig.5 represents  the dependence on $\l/\a$ of the ratio between  the displacement of the energy minimum and the cavity size $x=a_{min}/R$ for the same range $100 < R < 1000 $, while Fig.6 shows the dependence on $\l/\a$ of the relative distance between  the boundary surface and the position of the minimum $1-x=(R-a_{min})/R$, which to a certain degree turns out to be more informative, than the relative displacement  itself (see discussion below given in connection with the asymptotics for $R \to \infty$).

 \begin{figure}
  \includegraphics[width=8.6 cm]{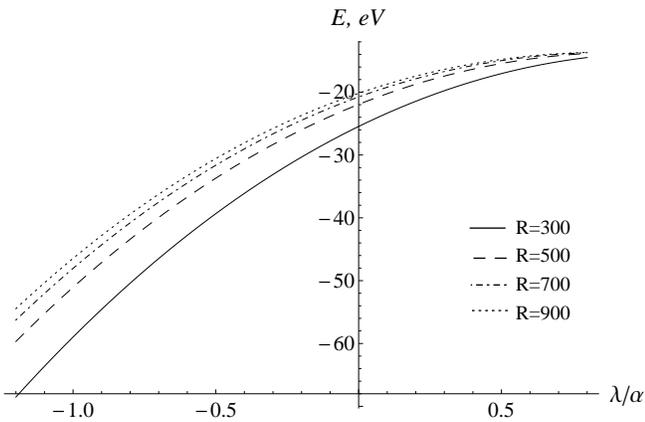}
\caption {The dependence of the shifted energy minimum $E_{min}=E_0(a_{min})$ on $\l/\a$.}
\label{fig:Emin_of_lambda}
\end{figure}

  \begin{figure}
  \includegraphics[width=8.6 cm]{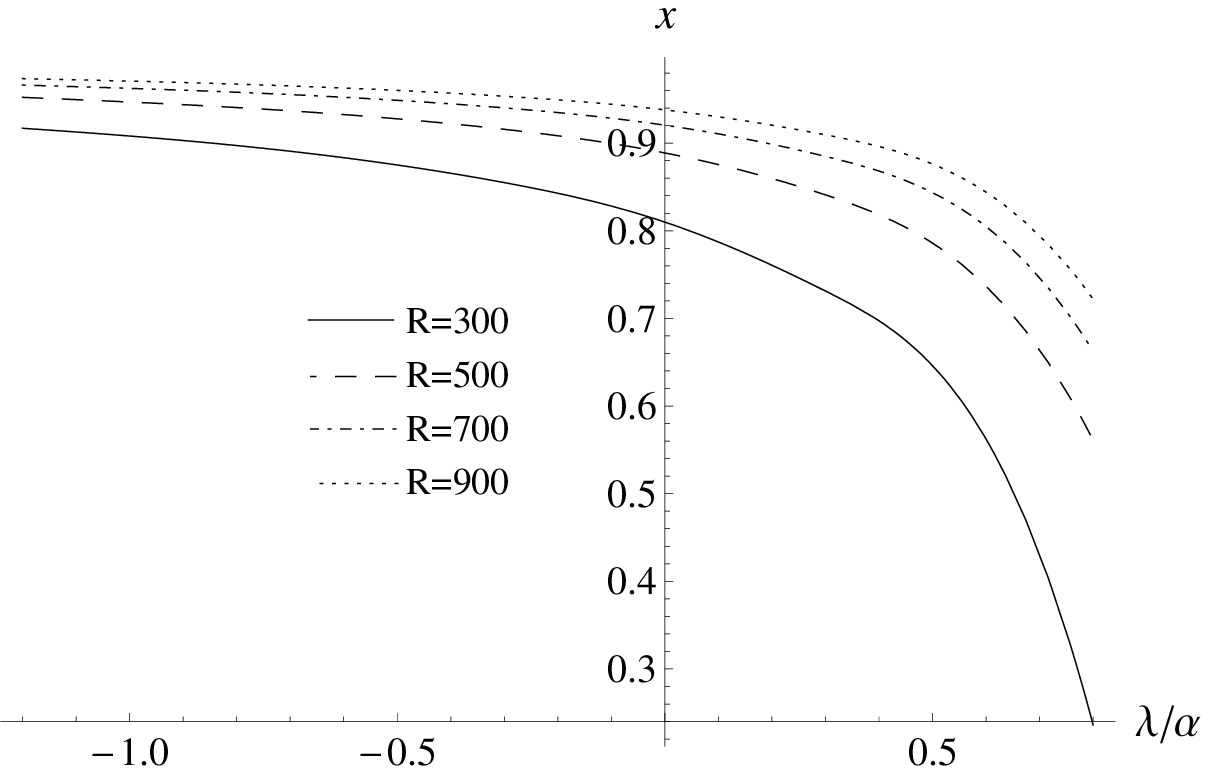}
\caption {The dependence of the parameter $x=a_{min}/R$ for the  energy minimum on $\l/\a$.}
\label{fig:xmin_of_lambda}
\end{figure}

  \begin{figure}
  \includegraphics[width=8.6 cm]{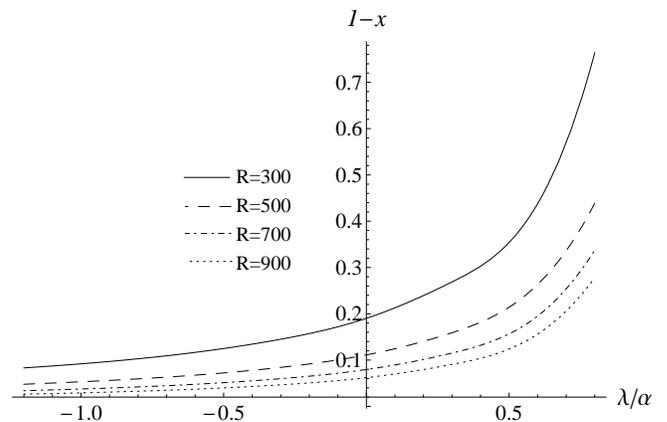}
\caption {The dependence of the parameter $1-x=(R-a_{min})/R$ for the energy minimum on $\l/\a$.}
\label{fig:one_minus_x_of_lambda}
\end{figure}

The dependence on  $R$ turns out to be more cumbersome, since  there appears now a specific scale of length in the problem --- the critical cavity radius  $R^{\star}(\l)$, which separates the regions of existence and absence of the displacement effect. Therefore in dependence on relation between $R$ and $R^{\star}$ the atomic behavior turns out sufficiently different. For $R \leq R^{\star}$ the displacement is absent, for $R \simeq R^{\star}$ the effect is extremely small, since the lowest level WF is dominated by the $s$-wave, and so there appears the scenario of an H in the cavity center, while the displacement shows up as a perturbation of the central problem only. For  $R >  R^{\star}$  angular harmonics with $l \not = 0$ start to produce a significant contribution, there appears a marked deformation of electronic WF and the displacement effect  grows rapidly. For $ R \gg  R^{\star}$ there appears another effect --- since the boundary curvature becomes small, the displacement tends to the asymptotical regime of  an H ``soaring'' over a plane at some height, defined by actual values of $q$ and $\l$.

By itself the problem of an H-like atom ``soaring'' over a plane with boundary condition (\ref{f3}) requires a separate and highly  nontrivial analysis, since in this case an interplay between two different symmetries --- spherical Coulomb and the axial one, caused by the boundary condition on a plane, takes place. Therefore the problem requires quite different methods of analysis and won't be discussed here in detail. Here only a brief sketch of most important features of this problem, which are necessary for correct interpretation of results for a box of extremely large size, will be presented. Firstly, for $\l \geq q$ a sufficiently strong reflection, which pushes H infinitely far from the plane, takes place. This is because for such $\l$ and any finite $R$ the atom resides in the cavity center.  It should be noted, that already this effect is nontrivial, since an infinite atomic jump away from the plane takes place for finite $\l \geq q$. On the contrary, for $\l < q$  the energy of the lowest electronic level is minimized at finite distances $d$ between them. This statement could be verified via following variational estimate. Let us pass to the cylindrical coordinates $(\r , \vf , z)$, when the lowest level WF takes the form $\p (\r , z)$, the position of atomic nuclei over a plane $z=0$ is given by  $\v d =(0,0,d)$, while the corresponding energy functional is written as
  \beq \label{f391}
E[\p]=\int_0^{\infty} \! dz \int \! \r d\r \ \left[  \1/2 | {\vec \nabla } \p|^2 + U(\r , z ) \ |\p|^2 \right]
\ + \nonumber \\ \eeq
\beq \label{f391}
+ \ { \l \over 2} \int \! \r d\r \  |\p(\r,0)|^2  \ ,
\eeq
where
 \beq \label{f3911}
 U(\r , z)=- q / \sqrt { \r^2 + (z-d)^2} \ .
\eeq
The pertinent trial function is chosen in the form
\beq \label{f392}
\P_{tr}(\r,z)=C \exp \[ -A \sqrt{\r^2+(z-d)^2} -B (z-d) \] \ ,
\eeq
with $C$ being the normalization factor, while $A,B$ are the variational parameters. To establish  the existence of a nontrivial minimum in $E_0(d)$ for finite $d$ and $\l<q$, it suffices  to deal with fixed values of parameters $A=1$,  $B=0$ without exploring the variational procedure for their definition. In this case   one obtains from   (\ref{f391}) for the estimate of the lowest level energy
\beq \label{f393}
E [\P_{tr}]=- {q^2 \over 2} {1- e^{-2 q d} (1+  2\l d  +\l/q -q d )/2 \over 1- e^{-2 q d} (1+ q d)/2 } \ .
\eeq
Now let  $ q d \gg 1$, what means actually that $d \gg a_B$. Then it is possible  to represent (\ref{f393}) as a series in powers of $\exp (-2 q d)$, what leads to the following result for $E_{tr}$
\beq \label{f394}
E_{tr}=- q^2 / 2 +  e^{-2 q d} \[ \l/q + 2 q d (\l/q -1)  \] + O ( e^{-4 q d}) \  .
\eeq
At this stage we recall the results  of the preceding section, that $E_0(d \to \infty)=-q^2/2 \ $ for $ \l  > -q $ and  $E_0(d \to \infty)=-\l^2/2 \ $ for $ \l < -q$. Proceeding further, from (\ref{f394}) one finds, that when $|\l/q| < 1$, there holds for sufficiently large $ q d $, that $E_{tr}< -q^2/2$, i.e. for the trial function of the form  (\ref{f392}) $E_{tr}$ turns out to be smaller, than the exact value of electronic energy for infinite distance between the atom and plane. So the minimal energy of the electronic level is achieved in this case for (possibly quite large), but definitely finite distance between the atom and plane. At the same time, for $\l < - q$  this effect should be even stronger, since the electronic attraction to the plane increases. The circumstance, that the function (\ref{f392}) doesn't satisfy the boundary condition (\ref{f3}), which takes now the form
\beq \label{f395}
\left. \(  \pd / \pd z  - \l  \) \p(\r,z)\right|_{z=0} =0 \ ,
\eeq
cannot pose any problems for  the status of the estimate considered above. The reason  is that  (\ref{f3}) and  (\ref{f395}) appear as additional equations within the variational problem for energy functionals (\ref{f1}, \ref{f391}), caused by restrictions on the integration region, and so should be satisfied only by the exact solution, corresponding to the true energy minimum, while $E_{tr}$, obtained from (\ref{f391}) via integration over the same region $z \geq 0$, turns out to be a correct estimate for  exact minimum of the functional (\ref{f391}) from above.

By means of  (\ref{f394}) it is also possible to estimate the distance from the plane via position of the minimum of $E_{tr}(d)$
\beq \label{f396}
d_{min}= {1 \over  2(1-\l/q) } \ .
\eeq
Here it should be noted once more, that formulae  (\ref{f394}) and (\ref{f396}) can be used as estimates for the  position of electronic energy minimum for sufficiently large $d$ and correspondingly $d_{min}$ only. Actually they could be remarkably different  from the true energy minimum, since the parametes $A$ and $B$ in the trial function (\ref{f392}) are chosen in the simplest way, ignoring the variational procedure, just  in order to establish the fact of attraction between the atom and plane for $\l < q$. Nethertheless, as the (quasi)exact numerical solution shows, these estimates agree quite well with the true minimum, whenever the latter is reached at sufficiently large distance between the atom and plane.

The results of numerical calculations of correct values for the electronic energy minimum of atomic H over a plane, obtained via gradient descent method, are shown in  Tab.2.
\beq {  \begin{array}{c|c|c}
\l/q  &  d_{min}  &  E_{min}  \\
 \hline
 0.8  &  324.7 & -13.61  \\
 0.6  &  158.6 & -13.90  \\
 0.3  &  86.0  & -15.45  \\
 0.0  &  57.3  & -18.47  \\
-0.3  &  42.3  & -23.20  \\
-0.6  &  33.4  & -29.68  \\
-1.2  &  23.3  & -48.72  \\
\end{array} }\notag \eeq
{\small Tab.2. (Quasi)exact values for the position and value (in eV) of the  energy  minimum for the lowest  electronic level of atomic H over a plane.} \vskip 0.3 cm
First line in Tab.2 shows, that the  minimum of electronic energy of atomic H  for $\l=0.8 \a $ lyes sufficiently far from the plane --- $d_{min}\simeq 2.37 a_B$, and so agrees quite well with estimate (\ref{f396}) presented above, which gives in this case for $d_{min}$  the value $\simeq 2.5 a_B$.

Now let us present (quasi)exact numerical results for the behavior of  atomic H in a cavity of large size $R$, which show explicitly, how it approaches the asymptotical regime of ``soaring'' over a plane. Figs. 7-9  demonstrate the dependence on $R$ of the bound energy and displacement for the lowest electronic level in the range  $-1.2 \leq \l/\a \leq 0.8$. It should be specially noted, that for $\l=0.8 \a \simeq\l^{\star}$  the displacement effect  presented on Figs. 7,8  shows up very weakly (see Fig.10 below too), since it takes place only in cavities with $R> R^{\star} \gg a_B$, when the displacement itself turns out to be quite large too, while the decrease of the level energy is very small. Therefore  for such  $\l$ an enhanced precision by calculation of the effect  is required, what is achieved via smooth extrapolation of contributions of angular harmonics from the range  20-30-40-50 to more large values, that allows in turn for a correct account of contribution from higher orbital momenta up  to $l \sim 100-200$ and even more.

On Fig. 7 the curves for the electronic bound energy as a function of $R$ are shown,
  \begin{figure}
  \includegraphics[width=8.6 cm]{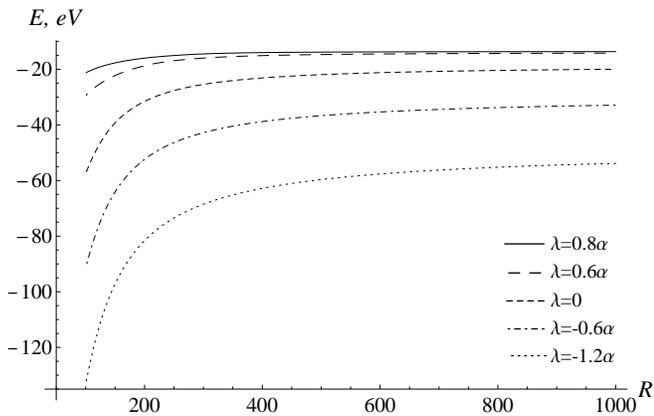}
\caption {The behavior of $E_{min}=E_0(a_{min})$ as a function of $R$ for fixed  $\l$ from the range $-1.2 \leq \l/\a \leq 0.8$.}
\label{fig:Emin_of_R_2}
\end{figure}
Fig.8 represents the dependence on $R$ for the relative displacement in units of $R$ of the energy minimum  $x=a_{min}/R$ for the same  $\l/\a$,
  \begin{figure}
  \includegraphics[width=8.6 cm]{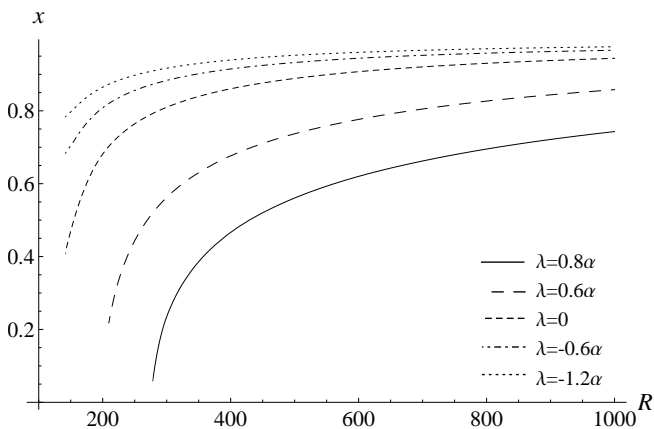}
\caption {The dependence of the parameter  $x=a_{min}/R$ for the electronic energy minimum on $R$}
\label{fig:xmin_of_R_2}
\end{figure}
while Fig.9 displays the distance between the box boundary and the position of the energy minimum $d_{min}=R-a_{min}$.
 \begin{figure}
  \includegraphics[width=8.6 cm]{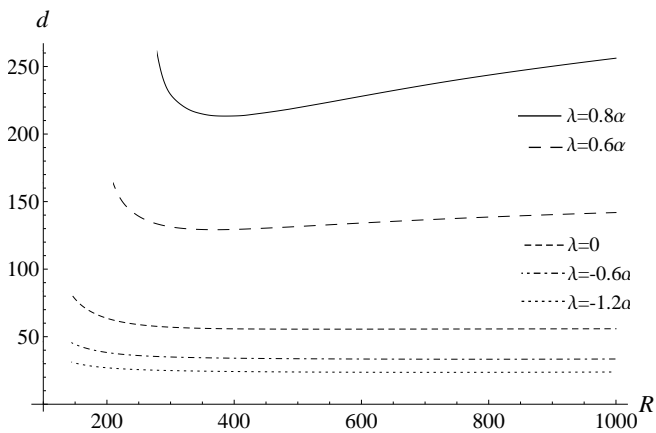}
\caption { The dependence of the distance between the cavity boundary and the electronic  energy minimum $d_{min}=R-a_{min}$ on $R$.}
\label{fig:R_minus_a_of_R_3}
\end{figure}
On the contrary to Figs. 7,8, each curve shown on Fig.9 contains a minimum for finite $R$ from the range $100 < R < 1000 $, which is well pronounced for $\l=0.8 \a \simeq\l^{\star}$, but for decreasing $\l$, hence for increasing attraction to the boundary, becomes rapidly very weak. The reason is that for such box sizes the distance between the atom and the box boundary approaches its asymptotical value corresponding to the atomic ``soaring'' over a plane  more and more quickly, while for $\l=0.8 \a \simeq\l^{\star}$ in the range $100 < R < 1000 $ this effect isn't present yet.

A comparison of results for a box of a large size $R= 7 a_B \simeq 1000$ and for a plane with $R=\infty$ is given in Tab.3, which confirms the result quoted above --- with growing $\l/q$ a more and more large $R$ is required for approaching the asymptotical regime.

\beq {  \begin{array}{c|c|c|c|c}
\l/q & d_{min}(\infty) & d_{min}(7 a_B)& E_{min}(\infty)& E_{min}(7 a_B)  \\
 \hline
 0.8 & 324.7 & 253.5  & -13.61 & -13.68  \\
 0.6 & 158.6 & 141.4  & -13.90 & -14.23  \\
 0.3 & 86.0  & 81.5   & -15.45 & -16.37  \\
 0.0 & 57.3  & 55.8   & -18.47 & -20.18  \\
-0.3 & 42.3  & 41.8   & -23.20 & -25.81  \\
-0.6 & 33.4  & 33.3   & -29.68 & -33.32  \\
-1.2 & 23.3  & 23.3   & -48.72 & -54.48  \\
\end{array} }\notag \eeq

{\small Tab.3. The position and magnitude (in eV) of the  energy minimum for the lowest electronic level of atomic H over a plane ($R=\infty$) and for a cavity of a large radius ($R= 7 a_B$)}.

\subsection*{5. Atomic H ground state shifted from the center of cavity}

Now let us turn to the dynamics of atomic H as a whole  by treating the position of atomic nuclei $\vec a$ as a dynamical variable. The dynamics of  nuclei restores the broken by atomic displacement initial $SO(3)$ via rotations of the shifted atom around the cavity center, whereas angular components of $\v a$ serve as Goldstone modes, which describe fluctuations  of spontaneous average (atomic displacement) under $SO(3)$ group transformations.

Within adiabatic approximation the corresponding effective hamiltonian for the dynamics of atomic nuclei takes the form
 \beq \label{f40}
H_{eff}=-{1 \over 2 M} \D_{\vec a} + E_0 (a) + E_{rec}(a) \ ,
\eeq
with $M$ being the nuclei (proton) mass, $E_0(a)$ is the lowest electronic level, considered in detail in the preceding section, while $E_{rec}(a)$ is a specific effect, analogous to the recoil effect for the free atom, when a correction
 \beq \label{f41}
\D E_{rec}= { m \over M} \< { {\vec p}^2 \over 2 m } \>_{el} \
\eeq
appears. In the case under consideration $E_{rec}(a)$ is caused by nuclei back-reaction  on the electronic WF deformation and takes the same form
 \beq \label{f42}
E_{rec}(a)= { 1 \over 2 M} \< \p_{el} | {\vec p}_a^2 | \p_{el} \>
\eeq
with that crucial difference, that now ${\vec p}_a=-i {\vec \nabla}_a$  doesn't possess the status of spatial translations generator for the electronic $\p_{el}(\vec r, \vec a)$, rather it  defines the ``kinetic'' effect of electronic WF distortion  caused by atomic displacement. The most consistent way to derive the expression (\ref{f42}) is based on field quantization in the vicinity of a boson soliton by means of collective (group) variables (see \cite{Raja}-\cite{Dub} and refs. therein). In the  case under consideration  the role of  bosonic soliton is played by the atomic nuclei with Coulomb field, while the electronic state appears as one-particle excitation in the Furry picture for fermion (electron-positron) field, that is considered in soliton background at the same footing with secondly quantized bosonic component. For a translationally invariant system, when the total momentum is conserved, the kinetic energy operator to the leading order of expansion in inverse powers of soliton mass takes the form \cite{Sv3},\cite{Dub}
 \beq \label{f43}
E_{kin}= { 1 \over 2 M} \[ \v P \ + \ :\int \! d\v r \ \v \nabla \F \ \Pi: +
:\int \! d\v r \ \c^+ \ i \v \nabla\c: \]^2 ,
\eeq
with $M$ being the soliton (atomic nuclei) mass, $\v P $ is the total momentum, while $\F , \Pi$ and $\c^+ , \c$ denote the secondly quantized boson field and its canonically conjugated momentum and the fermion field in the soliton rest frame. The normal ordering in (\ref{f43}) provides the validity of condition, that in the plane-wave basis for $\F$ and $\c$ the kinetic energy of one-particle state with momentum $\v k$ should be of the form
\beq \label{f44}
E_{kin}=  \( \v P \ -  \v k \)^2/ 2 M = {\v P}_{sol}^2/2 M \ ,
\eeq
corresponding to the kinetic soliton energy with recoil. For an atom in a box (\ref{f43}) transforms into
 \beq \label{f45}
E_{kin}= { 1 \over 2 M} \[i {\vec \nabla}_a \ + \ :\int \! d\v r \ \v \nabla_a \F \ \Pi: \ + \right.
\nonumber \\ \eeq
\beq \label{f45}
\left.  + \
:\int \! d\v r \ \c^+ \ i \v \nabla_a\c: \]^2 .
\eeq

Such a structure of the kinetic term could be easily verified from condition, that the transition into soliton (nuclei) rest frame should be a canonical transformation \cite{Sv3},\cite{Dub}. But ${\vec p}_a=-i {\vec \nabla}_a$ isn't conserved yet. For the bosonic component ${\vec \nabla}_a$  coincides up to the sign with ${\vec \nabla}_r$, since the deformation of atomic nuclei by displacement is negligibly small, even if it approaches the cavity boundary, thence soliton mass  in (\ref{f45}) remains the same as in (\ref{f43}). However, for the fermion field due to boundary condition (\ref{f3}) the dependence on  $\v a$ and $\v r$  becomes  by approaching the cavity boundary sufficiently diverse. Proceeding further,  in  nonrelativistic approximation for a one-particle fermion state, corresponding to the lowest atomic electron level, one finds  from (\ref{f45}) the ``recoil'' effect in the form (\ref{f42}),  since the lowest level electronic WF is real-valued  and so $\< \p_{el} | {\vec p}_a | \p_{el} \>=0$, whereas $\< \p_{el} | {\vec p}_a^2 | \p_{el} \> \not = 0$. At the same time,  the contribution of positron states (lower continuum of the Dirac equation) and the vacuum shift, which appear by taking average of operator (\ref{f45}) over one-electron valence state,  are dropped, since they have sense only by taking account of  relativistic corrections.

Now let us take into account, that in the spherical cavity $E_0(a)$ and $E_{rec}(a)$ are wittingly rotationally invariant, and so the initial  $SO(3)$ restores, the atomic state acquires rotational quantum numbers $JM_J$,   while the atomic energy levels in a cavity are defined from the radial equation
 \beq \label{f46}
\[ -{ 1 \over 2 M a^2} \pd_a (a^2 \pd_a) + E_0 (a)   \ + \right.
\nonumber \\ \eeq
\beq \label{f46}
\left. \ + { J (J+1) \over 2 M a^2 } + E_{rec}(a) \] \ \f_{at}
= E \ \f_{at} \ .
\eeq
In (\ref{f46}) due to large nuclei mass the orbital term  is important at the origin of coordinates only, while in $E_{rec}(a)$ the matrix element $\< \p_{el} | {\vec p}_a^2 | \p_{el} \>$  itself turns out to of the same order, as the electron kinetic energy and so $E_0(a)$, and shows up the same growth for $a \to R$. Moreover, in fact the angular part in $\< \p_{el} | {\vec p}_a^2 | \p_{el} \>$ coincides precisely with corresponding term in $\< \p_{el} | {\vec p}_r^2 | \p_{el} \>$. But due to the factor $1/2 M$ the recoil term $E_{rec}(a)$  turns out to be as small as the orbital one compared to $E_0(a)$. As a result, in (\ref{f46}) the leading contribution to $V_{eff}$ is given by the electronic energy $E_0(a)$. In turn,  due to kinetic energy of deformed electronic WF  there appears a potential barrier in $E_0(a)$ by $a \to R$ (see Fig.10), which  provides the  confinement of atomic nuclei inside the cavity via its large mass.
 \begin{figure}
  \includegraphics[width=8.6 cm]{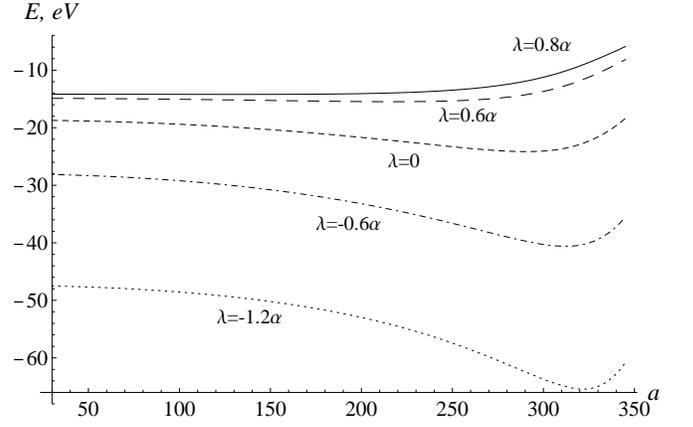}
\caption {$E_0(a) $ for $R=350, \ -1.2 \leq \l/\a \leq 0.8 $.}
\label{fig:barrier_R_s_1}
\end{figure}

Atomic ground state energy levels are given in Tab.4. As expected, they are shifted from the minimum of the effective potential $E_0(a)$ by several tenth of eV.

\beq {  \begin{array}{c|c|c}
\l/q & E_{at} & \D E_{at}  \\
 \hline
 0.8  &  -14.174  &  0.032 \\
 0.6  &  -15.363  &  0.11   \\
 0.3  &  -18.663  &  0.223  \\
 0.0  &  -23.799  &  0.345  \\
-0.3  &  -30.906  &  0.473  \\
-0.6  &  -39.986  &  0.604  \\
-1.2  &  -64.509  &  0.883  \\
\end{array} }\notag \eeq

{\small  Tab.4. The lowest atomic level $E_{at}$ and its shift from the minimum of effective potential $\D E_{at}=E_{at}-V_{min}$ in eV for $R=350$ and 7 values of  $\l/\a$ from the range  $-1.2\leq \l/\a \leq 0.8$. }\vskip 0.5 cm

Radial components of ground state atomic WF and density distribution  are shown on Fig. 11,12.
 \begin{figure}
  \includegraphics[width=8.6 cm]{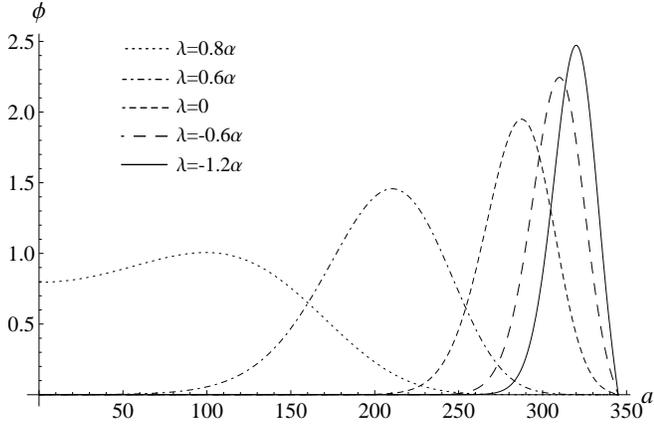}
\caption {Radial components of atomic nuclei ground state  WF for $R=350, \ -1.2 \leq \l/\a \leq 0.8 $.}
\label{fig:nuclear_wf}
\end{figure}
 \begin{figure}
  \includegraphics[width=8.6 cm]{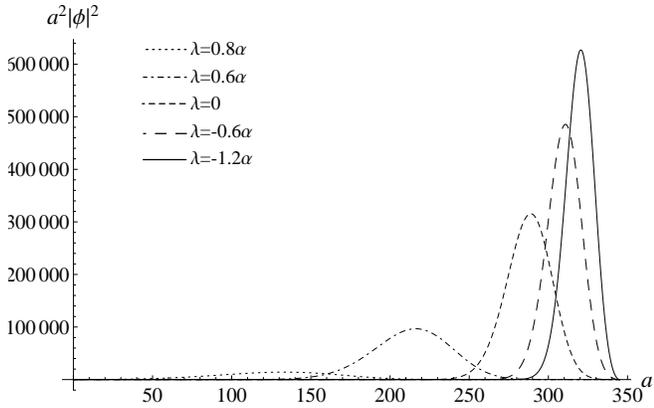}
\caption {Radial density distribution of atomic nuclei ground state for $R=350, \ -1.2 \leq \l/\a \leq 0.8 $.}
\label{fig:nuclear_dens}
\end{figure}
Note, that within the initial problem statement there are no special boundary conditions imposed on the atomic nuclei. The confinement of nuclei inside the  cavity volume proceeds dynamically by means of boundary condition  (\ref{f3}) for the electronic WF, whence the deformation of the latter works for $a \to R$ as a spring, returning the nuclei back into the cavity volume. Moreover, if $\l=0$, i.e. for Neumann boundary condition (\ref{f5}), which imply the possibility of periodic continuation of  WF through the box boundary, an electron might be spread over a (sub)lattice compiled from cavities, filled in by atoms of the same type, as it proceeds in the Wigner-Seitz model \cite{WS}. But each nuclei should be confined in its own cell within a spherical shell localized inside the cavity, as it is shown on Fig.12.

\subsection*{6. Conclusion }

So we have shown,   that the properties of such a ``not going out'' atomic state might in some cases be sufficiently different from the confinement by a potential barrier. The most remarkable feature here is, that  due to boundary condition (\ref{f3}) there come into play the nontrivial deformation properties of electronic WF under asymmetric distortion caused by interaction with the cavity boundary, whereas in the case of confinement by potential barrier the most essential role is played by  WF ``elasticity''  under uniform pressure \cite{Michels}-\cite{Aquino1}. In particular,  depending on the properties of the boundary the atomic position in the box center might turn out to be unstable and so the atom could be shifted towards the boundary. This displacement is accompanied with  increase of the electronic bound energy  and leads to  spontaneous breakdown of initial  $SO(3)$, thereon the Goldstone modes of atomic rotation restore the broken symmetry,  the stationary atomic states acquire quantum numbers $JM_J$ of the  total angular moment, while the atomic position  becomes spread over a spherical shell in the vicinity of the box boundary (see Fig.12). The confinement of atomic nuclei proceeds dynamically due to the boundary condition  (\ref{f3}) for the electronic WF --- the deformation of the latter works for $a \to R$ as a spring, which returns the nuclei back into the box volume.

As a consequence, the properties of  atomic H ground state turn out to be sufficiently different  in dependence on the cavity parameters. For definite values of $R$ and $\lambda$ the energy of the lowest level  could run  due to displacement effect up to values, that exceed the bound energy of the atom in the center of cavity by dozens of $eV$, while in the limit of very large $R$ the regime of an atom, soaring over a plane with boundary condition (\ref{f395}), is reproduced, rather than  a spherically symmetric configuration, what might be proposed on the basis of initial $SO(3)$ symmetry of the problem.

It should be noted also, that more complicated systems like molecular H ion behave in such a state with displacement towards the cavity boundary even more nontrivial, since due to boundary condition imposed on the electronic WF there changes  itself the mechanism of molecular formation.

To conclude let us emphasize once more, that the boundary condition (\ref{f3}) of ``not going out''  from the volume  $\W$ doesn't unavoidably imply an actual confinement of a particle inside $\W$, as it occures in the partial case of trapping by a potential barrier, but on the contrary, under definite conditions it allows for a much more wide problem statement, when a particle (electron) could be essentially delocalized \cite{Sv1}-\cite{Pupyshev1}. In particular, within Wigner-Seitz model \cite{WS} such an approach leads to a consistent description of the lowest particle state in a cubic (sub)lattice, formed by microcages of the same type inside the parental crystallic matrix. In the latter case instead of single energy level there appears the whole set of states $\p_{\v k}(\v r)=u_{\v k} (\v r) \exp (\ i \v k \v r)$ with condition $u_{\v k} (\v r +\v b )=u_{\v k} (\v r)$, where $\v b $ is the period of cavities sublattice, the wavevectors $\v k$ fill in the first Brilluen zone, while the periodic WF corresponds to the state with $\v k =0$ and defines position of the zone bottom. A concrete example of such kind is given by the considered in sect. 4,5 case of atomic H for  $\l=0$, which describes situation, when all the cavities,  occupied by atoms, form a  (sub)lattice similar to that of an alkaline metal.

The authors are very indebted to Dr. A.M.Puchkov from Inst. of Physics, St-Petersburg University, for drawing our attention to ref. \cite{ZRP}.

\end{document}